\documentclass[twocolumn,superscriptaddress,showpacs,aps,amsmath,amssymb,pra]{revtex4}

\begin{document}

\title{Lasing at arbitrary frequencies with atoms with broken inversion symmetry 
and an engineered electromagnetic environment}

\author{Michael Marthaler}
\affiliation {Institut f\"ur Theoretische Festk\"orperphysik,
      Karlsruhe Institute of Technology, 76131 Karlsruhe, Germany}

\author{Martin Koppenh\"ofer}
\affiliation {Institut f\"ur Theoretische Festk\"orperphysik,
      Karlsruhe Institute of Technology, 76131 Karlsruhe, Germany}

    \author{Karolina S\l{}owik}
\affiliation {Institut f\"ur Theoretische Festk\"orperphysik,
      Karlsruhe Institute of Technology, 76131 Karlsruhe, Germany}
\affiliation {Instytut Fizyki, Uniwersytet Miko\l{}aja Kopernika, Toru\'{n}, Poland}

\author{Carsten Rockstuhl}
\affiliation {Institut f\"ur Theoretische Festk\"orperphysik,
     Karlsruhe Institute of Technology, 76131 Karlsruhe, Germany}
\affiliation {Institute of Nanotechnology,
     Karlsruhe Institute of Technology, 76021 Karlsruhe, Germany}

\date{\today}

\begin{abstract}
With the purpose to devise a novel lasing scheme, we consider a two level system with both a transversal and longitudinal coupling
to the electromagnetic field. If the longitudinal coupling is sufficiently strong,
multi-photon transitions become possible. 
We assume furthermore that the electromagnetic environment has a spectrum
with a single sharp resonance, which serves as a
lasing cavity. Additionally, the electromagnetic environment should have a
very broad resonance around a frequency which differs form the sharp resonance. 
We use the polaron transformation and derive a rate equation to describe the dynamics of such system.
We find that lasing at the frequency of the sharp mode is possible, if the energy difference of the atomic
transition is similar to the sum of the frequencies of both peaks in the spectral function. This allows for
the creation of lasing over a large frequency range and may in perspective enable THz lasing at room temperature.
\end{abstract}

\maketitle

\section{Introduction}\label{sec:intro}

The standard source of coherent light from infrared to visible frequencies is the laser \cite{Review_scully}. 
A laser requires, first, population inversion in an active medium to create photons and, second, a cavity to allow for stimulated emission \cite{Book_scully}.
For an active medium consisting of natural atoms, all non-diagonal matrix elements of the dipole coupling between electromagnetic field and atom vanish 
because of the inversion symmetry of the atomic Coulomb potential \cite{QuantumDotswithoutPointSymmetry}.

However, artificial atoms with broken inversion symmetries, like 
 superconducting qubits \cite{Metamaterial_Pascal}, quantum dots \cite{Childress-PRA-69-042302,Doube_Dot_Jinshuang}, or molecules \cite{Meath-MolPhys51-3-585} have been studied as well. 
For such artificial atoms the typical Jaynes-Cummings Hamiltonian contains an additional $\sigma_z$-type coupling, which is often
 called longitudinal coupling in contrast to the standard 
 transversal $\sigma_x$ coupling. 
Lasing devices based on solid state qubits have been studied \cite{Doube_Dot_Jinshuang,Armour-SSET-Micromaser,Hauss_Lasing,Exp_Qubit_Lasing} and a large body of work exists on lasing with 
self-assembled quantum dots \cite{Dots_Laser1,Dots_Laser2,Dots_Laser3,Dots_Laser_4} and lasing with organic molecules \cite{MolecularLaser}. 
For superconducting devices, non-linearities based on the Josephson effect have been proposed as a way to create a non-linear coupling to the electromagnetic field
 \cite{Marthaler_Squeezed,Squeezed_Biased_Junction,Armour-SSET-Micromaser}.
In experimentally realized lasing devices, using superconducting qubits \cite{Astafiev-Nat-449-588,Chen-PhysRevB.90.020506,Dressed_State_Lasing_Jena}
or gate defined double dots \cite{Doube_dot_Laser_Petta},
the $\sigma_z$ coupling between artificial atom and cavity field is present but has not yet been studied. 

Artificial atoms, which have an additional longitudinal coupling to the electromagnetic field, can 
show a particularly rich behavior if they couple to a sharp cavity mode and additionally to a broad spectral density, which can be treated as a 
reservoir in equilibrium \cite{lasing_with_strong_noise_marthaler}.
The longitudinal coupling allows then for resonant multi-photon transitions.  
It has been shown that, depending on the spectral function of the bath, effects exist, which can enhance  
stimulated emission of such an artificial atom \cite{inversionless_petta,inversionless_marthaler}.
In this work, we consider a situation where an artificial two-level atom is coupled to an electromagnetic environment 
with an adjustable and well controlled spectral density.

A particular example would be to couple an artificial atom to a metallic nanostructure. These metallic nanostructures 
respond resonantly to light at different frequencies, where they may sustain localized surface plasmon polaritons \cite{Plasmon1}. 
Plasmonic nanostructures, therefore, can be used to tailor the local density of states of the electromagnetic modes \cite{Tam,Tame,Pelton}. 
In principle, it is possible to create a plasmonic structure with a relatively sharp mode and an additional spectrally broad,
dissipative contribution. This requires the use of a plasmonic nanostructure possessing, e.g. a trapped or a Fano type resonance \cite{Fano1,Fano2},
and a broad continuum with an electric dipolar response. This is the kind of structure we have in the following in mind.
We want the dissipative contribution in the spectral density to have a maximum at a frequency, which is the difference
between the level splitting of the artificial atom and the frequency of the sharp peak. We will show that
in this case it is possible to generate lasing at the frequency of the sharp mode. By flexibly tailoring 
the resonance frequency of the electromagnetic environment, this allows for lasing at a large range of frequencies. 

If the upper level of the two level system is excited beyond the thermal occupation, lasing becomes possible through the 
following process: The  longitudinal coupling allows a multiphoton process, a photon is created in both, the sharp mode and in the 
broad part of the spectral density. The energy for this process comes from the two level system, which relaxes into the ground state. 
The sum of the energies of the two photons has to be similar to the energy of the two level system. The excitation of the broad spectral density
can be assumed to relax fast, while the photon in the sharp mode has a long live time. This process creates an effective population inversion between the
state of the system with a photon in the broad spectral density and the state without this photon. This initiates the lasing process. The description of this entire process is at the heart of this publication.

To do so, we use in the following the polaron transformation to bring the Hamiltonian that describes the entire system into a form, which allows us to 
understand multi-photon transitions directly by looking at possible transition matrix elements.
The broad background spectrum can be used to dress the coupling between the sharp mode and the two level system in a way
that allows for a convenient calculation of transition rates. From this we can understand the properties of photon emission and absorption into and from the sharp mode.

The energy distribution of the electromagnetic modes is described by the Hamiltonians $H_{s}$ for the sharp mode and
$H_{d}$ for the broad mode spectrum,
\begin{eqnarray}
 H_{s} &=& \Omega a^{\dag}a\, ,\\
 H_{d}   &=& \sum_{k}\nu_k b_{k}^{\dag} b_k\, .\nonumber
\end{eqnarray}
The broadening of the sharp mode will be included using a standard master equation approach.

The annihilation (creation) operator of the sharp mode is given by $a$ ($a^{\dag}$). The broad spectrum is described by 
 the annihilation (creation) operators $b_k$ ($b_k^{\dag}$), which annihilate (create) an excitation in mode $k$. 

The considered two level system is characterized by a level splitting $\Delta E$. It is coupled to the 
sharp mode with the strength $g$, and additionally to each of the modes $k$ in the
broad spectrum with the strength $g_k$. We define the angle $\theta$ such that the transversal coupling is proportional to $\sin\theta$ and the longitudinal coupling is proportional to 
$\cos \theta$. To write the 
total Hamiltonian in a compact form, we define a 
quasi-coordinate $x$ and quasi-momentum $p$ for the sharp mode
\begin{eqnarray}
 x     &=& g\sin\theta(a^{\dag}+a)\, ,\\
 p     &=& i\frac{g\cos\theta}{\Omega}(a^{\dag}-a)\, .\nonumber 
 \end{eqnarray}

Similarly for the broad spectrum we define 
 a quasi-coordinate $\bar{x}$ and quasi-momentum $\bar{p}$
 \begin{eqnarray}
 \bar{x} &=& \sin\theta\sum_k g_k (b_k^{\dag}+b_k)\, ,\\
 \bar{p} &=& i\cos\theta\sum_k \frac{g_k}{\nu_k}(b_{k}^{\dag}-b_k)\, .\nonumber
\end{eqnarray}

In the next section we will discuss the transformation of the basic Hamiltonian
of our system. Then we will discuss how to derive transition rates
between the eigenstate of the system.
In the last section we will discuss how to create lasing within our model.

\section{the system}
We will consider a two level system, 
coupled to a single sharp mode and to an additional reservoir
\begin{eqnarray}
 H=\frac{1}{2}\Delta E \sigma_z +(\sigma_x+\sigma_z \cot \theta ) (x+\bar{x})+H_s + H_d.
\end{eqnarray}
Here, $\sigma_i$ are the Pauli matrices acting on the artificial atom. The size of the 
longitudinal dipole matrix element is characterized by the angle $\theta$.
We can transform this Hamiltonian using the polaron transformation
\begin{equation}
 U=e^{i\sigma_z(p+\bar{p})}\,.
\end{equation}
After the unitary transformation we have
\begin{eqnarray}
 H_{P}      &=& U^{\dag} H U\nonumber\\
            &=& H_0 + H_g + H_d, \nonumber\\
 H_0        &=& \frac{1}{2}\Delta E \sigma_z+\Omega a^{\dag}a,       \label{eq_system_Hamiltonian}\\
 H_g        &=& \left(\sigma_+ e^{-i2\bar{p}} e^{-ip}x e^{-ip} +{\rm h.c.}\right),\label{eq_coupling_Hamiltonian_Photon_Creation} \\
 H_d        &=& (\sigma_+ e^{-i2p} e^{-i\bar{p}}\bar{x}e^{-i\bar{p}} + {\rm h.c.} ). \label{eq_coupling_Hamiltonian_Decay}
            \end{eqnarray}
Here, we have divided the Hamiltonian in a part which contains the system
(\ref{eq_system_Hamiltonian}),
and two parts which contain the coupling between system and reservoir, (\ref{eq_coupling_Hamiltonian_Photon_Creation}) and  (\ref{eq_coupling_Hamiltonian_Decay}) respectively.
We will treat the coupling perturbatively in the lowest order. The coupling terms 
create transitions between the eigenstates of the system, which consists of the two level system, described by $\Delta E\sigma_z/2$ and the 
sharp mode described by $\Omega a^{\dag}a$.
The eigenstates are given by the product base
of the two level system and the photon states of the sharp mode, $|o,n\rangle$.
The variable $o$ can take the values $+$ and $-$ with $\sigma_z|\pm\rangle=\pm|\pm\rangle$.
The photon states are numbered by $n$, with $a^{\dag}a|n\rangle=n|n\rangle$.
Therefore, the states have the energy
$(\Delta E\sigma_z/2+\Omega a^{\dag}a)|\pm,n\rangle=(\pm\Delta E/2 + \Omega n)|\pm,n\rangle$.

The transition rate from the state 
$|o,n\rangle$ to the state $|o',m\rangle$ is given by
\begin{equation}
 \Gamma_{o,n\rightarrow o',m}=
 \Gamma_{o,n\rightarrow o',m}^{\rm decay}+\Gamma_{o,n\rightarrow o',m}^{\rm ph}\,, 
\end{equation}
where we divided the rate into two components. The first rate $\Gamma_{o,n\rightarrow o',m}^{\rm decay}$ 
corresponds to an expansion of the term (\ref{eq_coupling_Hamiltonian_Decay}) 
and the second rate $\Gamma_{o,n\rightarrow o',m}^{\rm ph}$ corresponds 
to an expansion of (\ref{eq_coupling_Hamiltonian_Photon_Creation}), respectively. 

The physical meaning of these two rates is well defined for $\theta=\pi/2$, i.e. for a purely transversal coulping. In this case we have
$\Gamma_{o,n\rightarrow o',m}^{\rm decay}\propto \delta_{n,m}$. This means $\Gamma_{o,n\rightarrow o',m}^{\rm decay}$ corresponds
to a decay rate of the two level system,
but leaves the photon number in the sharp mode unchanged. For $\theta=\pi/2$ the rate $\Gamma_{o,n\rightarrow o',m}^{\rm ph}$ obeys the condition
\begin{eqnarray}
 \Gamma_{+,n\rightarrow -,m}^{\rm ph}\propto \delta_{m,n+1}\,\,\, , \,\,\, \Gamma_{-,n\rightarrow +,m}^{\rm ph}\propto \delta_{m,n-1}~.  
 \end{eqnarray}
This means $\Gamma_{o,n\rightarrow o',m}^{\rm ph}$
contains the transition \mbox{$\uparrow,n\rightarrow\downarrow,n+1$}, which is the transfer of an excitation of the two 
level system into the photon field which is crucial for lasing.  

In the following we will discuss the relevant noise correlators, which are connected to the transition rates.
To calculate $\Gamma_{o,n\rightarrow o',m}^{\rm ph}$ we need an explicit form for the correlator $\langle e^{-2i\bar{p}}(t)e^{2i\bar{p}}(0)\rangle$
while for $\Gamma_{o,n\rightarrow o',m}^{\rm decay}$ we need $\langle (e^{-i\bar{p}} \bar{x} e^{-i\bar{p}})(t)(e^{i\bar{p}} \bar{x} e^{i\bar{p}})(0)\rangle$. With these results at hand we will calculate the transition rates and show that they correspond to lasing.

\subsection{Polaronic coupling to the reservoir}

If a system couples to a reservoir via an operator of the form
$ e^{-2i\bar{p}}$, we have to consider the noise correlator
\begin{equation}
 C_P(t)=\langle e^{-2i\bar{p}}(t)e^{2i\bar{p}}(0)\rangle,
\end{equation}
where the averaging is performed by tracing over the bath degrees of freedom. 
This correlator is well known \cite{Nazaron_pofe} and can be written in an explicit form as
\begin{equation}\label{eq_P0fE_In_time_space_expclicit}
 C_P(t)=e^{\frac{4}{\pi}\int_0^{\infty} d\omega\frac{J(\omega)}{\omega^2}\left[\coth\frac{\omega}{2k_B T}(1-\cos\omega t)-i\sin\omega t\right] },
\end{equation}
where $J(\omega)$ is the spectral density of the reservoir. 
We will later discuss in more detail how this correlator corresponds to transition rates in the system. 
The spectral function can be connected to the effective impedance $Z(\omega)$ of the reservoir by
\begin{equation}
 J(\omega)=2\pi\omega\, {\rm Re}\, Z(\omega)/R_K\, ,
\end{equation}
where $R_K=h/e^2$ is the resistance quantum. 

The impedance, which we want to consider, is the impedance of an electromagnetic mode with frequency 
$\omega_L$. We want this mode to be very broad, such that we can assume that it always stays in equilibrium. However, we will first recall
the result for a perfectly sharp mode and we will later introduce the broadening phenomenologically. The impedance for a sharp mode is given by
\begin{equation}\label{eq_Impedance_for_Resonator}
 \frac{2{\rm Re}\,Z(\omega)}{R_K}=\epsilon_C\left(\delta(\omega-\omega_L) +\delta(\omega+\omega_L) \right),
\end{equation}
where $\epsilon_C$ is a coupling constant. In this very particular case given by Eq.~(\ref{eq_Impedance_for_Resonator})
this would correspond to $\epsilon_C=g_k^2/\nu_k$, where the wavevector $k$ is fixed by the relationship to the frequency $\omega_L=\nu_k$.

For this example the explicit form of Eq. (\ref{eq_P0fE_In_time_space_expclicit}) is known and can be written as \cite{Nazaron_pofe}
\begin{eqnarray}
 C_P(t) &=& e^{-(\eta_+ + \eta_-)} \sum_{m,n}\frac{\eta_+^m\eta_-^n}{m!n!}e^{-i(n-m)\omega_L t}\\
      & &\eta_{\pm}=\frac{4\epsilon_C\cos^2\theta}{\omega_L}\frac{\pm 1}{e^{\pm \omega_L/k_B T}-1}~.\nonumber
\end{eqnarray}

To model the mode not as a sharp feature but to give it a certain width $\Gamma$, we now introduce 
the broadened spectral function
\begin{equation}
S_P(\omega)=\int_{-\infty}^{\infty}dt C_P(t) e^{i\omega t}e^{-\Gamma |t|}\, . 
\end{equation}
We will discuss later the various physical effects that can be part of $\Gamma$.
  
Using our definition for $S_P(\omega)$, the spectral function becomes 
 \begin{eqnarray} \label{eq_Sgforsinglemode}
  S_{\rm P}(\omega) &=& \exp\left(-[\eta_+ +\eta_-]\right)\\
   & &\times\sum_{n,m}\frac{\eta_+^m\eta_-^n}{n! m!}
  \frac{2\Gamma}{\Gamma^2+(\omega-(n-m)\omega_L)^2}\nonumber\, .\nonumber
 \end{eqnarray}
 The spectral function has many peaks. Each peak corresponds to 
 a process where the two level system changes its state, emits or absorbs a photon to/from the sharp mode, and 
 an additional $m$ photons to/from the broad spectral density.  We will later discuss particular features which appear for a resonance condition,
 $\Delta E-\Omega\approx\omega_L$.

\subsection{Broadened linear coupling to the reservoir}

If a system couples to a reservoir via an operator of the form
$ e^{-i\bar{p}} \bar{x} e^{-i\bar{p}} $, we have to consider the noise correlator
\begin{equation}
 C_D(t)= \langle (e^{-i\bar{p}} \bar{x} e^{-i\bar{p}})(t)(e^{i\bar{p}} \bar{x} e^{i\bar{p}})(0)\rangle.
\end{equation}
 To find an explicit form for this correlator, we use the
 generating function
 \begin{eqnarray}
  F=\langle T e^{i \int dt' \left(\mu(t')\bar{x}(t')-i\lambda (t')  \bar{p}(t')\right)}\rangle\, ,
 \end{eqnarray}
 where $T$ is the time sorting operator.
  Using this function, we can show that the relaxation correlator is given by
  \begin{eqnarray}\label{eq_apA_Cchfromgenerator}
  C_{\rm D}(t)=-\left. \frac{d^2 F}{d\mu(0)d\mu(t)}
    \right\rvert_{{\tiny
    \begin{array}{rl}
     \lambda(t')&=(\delta(t-t'+\eta)+\delta(t-t'-\eta))\\
                &-(\delta(t'+\eta)+\delta(t'+\eta))\\
     \mu(t')&=0
    \end{array}}}\, .\nonumber\\
  \end{eqnarray}
  The explicit form of the generating function reads
 \begin{eqnarray}
   F &=& \exp\left[-\int dt_1 \int dt_2 \tilde{F}(t_1,t_2)\Theta(t_1-t_2)\right]\, ,\\
   \tilde{F} &=&\left(\mu(t_1)\mu(t_2)\langle \bar{x}(t_1)\bar{x}(t_2)\rangle\right.\nonumber\\
           & & -i\mu(t_1)\lambda(t_2)\langle \bar{x}(t_1)\bar{p}(t_2)\rangle\nonumber\\
     & & -i\lambda(t_1)\mu(t_2)\langle \bar{p}(t_1)\bar{x}(t_2)\rangle\nonumber\\
   & & \left.   -\lambda(t_1)\lambda(t_2)\langle \bar{p}(t_1)\bar{p}(t_2)\rangle 
\right)\, .\nonumber
 \end{eqnarray}
 We can apply Eq.~(\ref{eq_apA_Cchfromgenerator}) to this form of the generating function and get
 \begin{eqnarray}
  C_{\rm D}(t)&=&\left[  \langle \bar{x}(t)\bar{x}(0) \rangle
                   +4\langle \bar{p}(t)\bar{x}(0) \rangle^2 \right]\\
           & &\times \langle e^{-2i\bar{p}(t)} e^{2i\bar{p}(0)} \rangle \, .\nonumber
 \end{eqnarray}
 
We now wish to consider again a situation with an impedance 
given by Eq. (\ref{eq_Impedance_for_Resonator}).
As before, we do not want a sharp mode, but it should have a certain width $\Gamma$.
Therefore, we define the spectral function
\begin{equation}\label{eq_Boradened_spectral_function}
S_D(\omega)=\int_{-\infty}^{\infty}\,dt\, C_D(t) e^{i\omega t}e^{-\Gamma |t|}\, . 
\end{equation}

To write the resulting spectral function in a compact form we first define
 \begin{eqnarray}
  S_{\rm D,eff}(\omega)&=&\frac{ \Gamma \omega_L^2\tan^2\theta}{2}\left(
 \frac{\eta_-}{\Gamma^2+(\omega-\omega_L)^2}\right.\\
   &+&\frac{\eta_+}{\Gamma^2+(\omega+\omega_L)^2}
    +\frac{\eta_+^2}{\Gamma^2+(\omega+2\omega_L)^2}\nonumber\\
   &+&\left.\frac{\eta_-^2}{\Gamma^2+(\omega-2\omega_L)^2}
    -\frac{2\eta_-\eta_+}{\Gamma^2+\omega^2}\right)\, .\nonumber
 \end{eqnarray}
 This spectral function together with Eq. (\ref{eq_Sgforsinglemode}),
 gives us the relaxation spectral function 
 \begin{eqnarray}\label{eq_Schforsinglemode}
  S_{\rm D}(\omega)&=& \exp\left(-[\eta_+ +\eta_-]\right)\\
   & &\times\sum_{n,m}\frac{\eta_+^m\eta_-^n}{n! m!}
   S_{\rm D,eff}(\omega-(n-m)\omega_L)\nonumber\, .
 \end{eqnarray}
  Here we see that the total spectral density $S_{\rm D}$
  consists of multiple peaks, which repeat with distance $\omega_L$.  
  This means relaxation is always strong for $\Delta E=\bar{m}\omega_L$, 
 with $\bar{m}=\pm1,\pm2,\ldots$.

\subsection{The transition rates}

We can now write down the transition rates as calculated by a second order expansion of $H_g$ and $H_d$.
The photon creation rate is given by
\begin{eqnarray}\label{eq_Photon_Creation_Rate_Version_One}
 \Gamma_{\pm,n\rightarrow \mp,m}^{\rm ph}&=&|\langle \mp,m|\sigma_{\mp} e^{\pm ip} x e^{\pm ip}|\pm,n\rangle|^2 \\
                                &\times & \int_{-\infty}^{\infty} 
 dt C_P(t) e^{i((n-m)\Omega\pm \Delta E)t}e^{-\Gamma|t|}\nonumber\\
  &=&|\langle \mp,m|\sigma_{\mp} e^{\pm ip} x e^{\pm ip}|\pm,n\rangle|^2 \nonumber \\ 
&\times & S_P((n-m)\Omega\pm \Delta E)\, .\nonumber
\end{eqnarray}
The decay rate can be written as
\begin{eqnarray}\label{eq_DEcay_rate_not_explicit}
 \Gamma_{\pm,m\rightarrow \mp,n}^{\rm decay}&=&|\langle \mp,n|\sigma_{\mp} e^{\pm 2ip}|\pm,m\rangle|^2\\
 &\times& \!\!\!\!\int_{-\infty}^{\infty} 
 dt C_D(t)
 e^{i((n-m)\Omega\pm \Delta E)t}e^{-\Gamma|t|}\nonumber\\
  &=&|\langle \mp,m|\sigma_{\mp} e^{\pm 2ip}|\pm,n\rangle|^2 S_D((n-m)\Omega\pm \Delta E).\nonumber
\end{eqnarray}
We want to point out that in contrast to standard golden rule rates, we have here the additional broadening $\Gamma$. As discussed above, it allows us 
to model phenomenologically the width of the dissipative contribution of the electromagnetic environment, while we can still use the formal analytical result, which can be derived for a 
sharp mode. 

But our rates still have the standard form of a transition matrix element multiplied by the spectral function evaluated at the relevant frequency.
An interesting fact about this transition matrix elements  is that they allow for  multi-photon
transitions. However, if a resonance condition applies,
which favors the transition which only changes the photon number by one photon, this will of course suppress all other transitions. 
Additionally, the matrix elements   $ \langle n| e^{\pm ip} x e^{\pm ip}|n+m\rangle  $  and  $ \langle n| e^{\pm 2 ip} |n+m\rangle  $
scale like $ (g\cos\theta/\Omega)^m$ for $g\cos\theta/\Omega \ll 1$. 
Therefore, we will consider the approximations
\begin{eqnarray}\label{eq_Approximation_for_matrix_elements}
 & & \langle n | e^{\pm 2ip}|n\rangle \approx 1\,\,\, ,\,\,\,\langle n | e^{\pm 2 ip}|n+m\rangle\approx 0 ,\\
 & & \langle n-1| e^{\pm ip}x e^{\pm ip}|n\rangle \approx g\sin^2\theta \sqrt{n}~, \nonumber\\
  & & \langle n| e^{\pm ip}x e^{\pm ip}|n-1\rangle \approx g\sin^2\theta \sqrt{n}~. \nonumber
\end{eqnarray}
Using this approximation we can simplify the rates to
\begin{eqnarray}\label{eq_rates_simplified_by_simplified_matrix_elements}
 \Gamma_{+,n\rightarrow -,n+1}^{\rm ph}&=&  \Gamma_{+,n}=   g^2 (n+1)\sin^2\theta S_P(\delta\omega)~,\\
 \Gamma_{-,n\rightarrow +,n-1}^{\rm ph}&=&  \Gamma_{-,n}=   g^2 n\sin^2\theta S_P(-\delta\omega)~,\nonumber\\
 \Gamma_{\pm,n\rightarrow \mp,n}^{\rm decay}&=& S_D(\pm\Delta E)~,\nonumber
\end{eqnarray}
where we have introduced the detuning $\delta \omega=\Delta E-\Omega$.

\subsection{Self-consistent photon creation rates}\label{subsec_Photon_Creation}

In the previous section we discussed the transition rates in the lowest order approximation.
In this section we want to go somewhat further. We will focus on the photon creation rate.
As can be seen by our approximation shown in Eq. (\ref{eq_Approximation_for_matrix_elements}), 
in general the transition matrix element for transitions that increase or decrease the number of 
photons by one, grows with the total number of photons. This is the effect known as stimulated emission (absorption),
which is a key ingredient of the lasing process. However, calculating a rate
in the lowest order is only possible as long as the relevant noise correlator has a decay rate larger then the 
prefactor of the rate \cite{Karlewski_Non_Markovian,Marthaler_Strong_Noise}.
It seems clear that this condition will be violated for larger photon numbers. Therefore,
we will use a self-consistent approach which will guarantee convergence for our rates.

We can make the rate defined in  Eq. (\ref{eq_Photon_Creation_Rate_Version_One})
a self-consistent equation, by including the rate itself as an effective broadening of the 
energy,
 \begin{eqnarray}\label{eq_Photon_Creation_Rate_Version_Selfconsistend_For_Single_Photon}
 \Gamma_{+,n}^s &=  &g^2 (n+1)\sin^2\theta    \int_{-\infty}^{\infty}  \!\!\!\!\!
  dt C_P(t)
 e^{i\delta\omega t} 
    e^{-\Gamma_T |t|/2} e^{-\Gamma|t|}\nonumber \\
    \Gamma_{-,n}^s &=  &g^2 n\sin^2\theta    \int_{-\infty}^{\infty}  \!\!\!\!\!
  dt C_P(t)
 e^{-i\delta\omega t} 
    e^{-\Gamma_T |t|/2} e^{-\Gamma|t|} 
\end{eqnarray}
with $\Gamma_T=\Gamma_{+,n}^s+ \Gamma_{-,n}^s$
and the detuning $\delta\omega=\Delta E-\Omega $. 
Additionally, instead of solving the full-self-consistent
equation, we approximate the rates in the exponents by the lowest order results,
$\Gamma_T=\Gamma_{+,n}+ \Gamma_{-,n}$. 
The selfconsistent approach used here has previously been discussed in relation to 
decoherence rates of superconducting qubits \cite{Catelani_qp,Zanker_qp}, transport across an Anderson quantum dot \cite{Anderson_quantum} 
and is often called self-consistent Born-approximation \cite{Self_consistent_born}.

We want to consider
the spectral function of a single mode
(\ref{eq_Impedance_for_Resonator}). If we calculate the rates
(\ref{eq_Photon_Creation_Rate_Version_Selfconsistend_For_Single_Photon}) for this spectral function, we get an infinite sum of Lorentzian functions
\begin{eqnarray}\label{eq_Photon_Creation_Rate_Version_Selfconsistend_For_Single_Photon_explicit}
 \Gamma_{+,n}^s &=  &g^2 (n+1)\sin^2\theta   
  e^{-(\eta_++\eta_-)}\\
   & & \sum_{m',n'}\frac{\eta_+^{m'}\eta_-^{n'}}{m'!n'!} \frac{2\Gamma+\Gamma_T}{(\Gamma+\Gamma_T/2)^2 + (\delta\omega-(n'-m')\omega_L)^2}~, \nonumber \\
    \Gamma_{-,n}^s &=  &g^2 n\sin^2\theta     e^{-(\eta_++\eta_-)}\\
   & & \sum_{m',n'}\frac{\eta_+^{m'}\eta_-^{n'}}{m'!n'!} \frac{2\Gamma+\Gamma_T}{(\Gamma+\Gamma_T/2)^2 + (\delta\omega+(n'-m')\omega_L)^2}~. \nonumber 
\end{eqnarray}
We will consider a case where we have $\delta\omega\approx \omega_L$. In this case, the photon creation rate $\Gamma_{+,n}^s $ is relatively large, independent of temperature. 
We see that the photon absorption rate $\Gamma_{-,n}^s$ is suppressed as long as $\omega_L \gg k_B T$. We want to emphasize that we only need $\omega_L\gg k_B T$ to create lasing, 
but $\Omega$, the frequency of the lasing cavity can be much smaller than temperature.

In the limit of large photon number $n$ we find
\begin{eqnarray}
  \Gamma_n^{{\rm ph},+} &\approx& \Gamma_n^{{\rm ph},-} \approx \frac{4}{S_P(-\delta\omega)+S_P(\delta\omega)}~.
\end{eqnarray}
Therefore, as we would expect for large photon numbers, absorption and emission are equal. 

\section{Lasing}
To find the density matrix in the stationary limit, we describe the system by a master equation. 
The master equation reads 
\begin{eqnarray}\label{eq:masterrhoincoherent}
\dot{\rho}
&=&
-i \, \left[\frac{1}{2}\Delta E \sigma_z +\Omega a^{\dag}a,\rho\right]\\
& &+
(L_{\rm g}+L_{\rm ch}+L_{\rm diss}) 
\, \rho 
\, , \nonumber
\end{eqnarray}
where the Lindblad operator is separated into thee parts. 
The coupling between oscillator and two level system is described by
\begin{eqnarray}\label{eq_Lindblad_Photon_creation}
L_{\rm g}  \rho
&=&
\sum_n \Gamma_{+,n}^{s} \left(
2 o_{n+1,n}\sigma_-\rho\sigma_+o_{n,n+1}\right.\nonumber\\
& & \left. - \sigma_+ o_{n,n}\sigma_-\rho
-\rho \sigma_+o_{n,n}\sigma_-
\right)
\nonumber\\ 
& &
+\sum_n \Gamma_{-,n}^{s} 
\left(
2\sigma_+ o_{n-1,n}\rho o_{n,n-1}\sigma_- \right.\nonumber \\
& &\left.- 
\sigma_-   o_{n,n}\sigma_+\rho 
- \rho  \sigma_- o_{n,n} \sigma_+  \right)
\,, \nonumber
\end{eqnarray}
where we introduced the operator $o_{n,m}=|n\rangle\langle m|$.
Excitation and relaxation of the two level system is contained in the superoperator $L_{\rm ch}$, which reads
\begin{eqnarray}\label{eq_Lindblad_qubit_decay}
L_{\rm ch} \, \rho
\!\! &=& \!\!
\frac{\Gamma_{+}}{2}
\left(2\sigma_+\rho\sigma_- - \sigma_- \sigma_+\rho
-\rho \sigma_- \sigma_+\right)
\nonumber\\ 
\!\! &+& \!\!
\frac{\Gamma_{-}}{2}
\left(2\sigma_- \rho \sigma_+ - 
\sigma_+ \sigma_- \rho-\rho \sigma_+ \sigma_- \right)
.~~
\end{eqnarray}
The excitation and relaxation rates are given by $\Gamma_{\pm}=\Gamma_{\pm}^0+\Gamma_{\mp,n\rightarrow\pm,n}^{\rm decay}$. 
Here, the rates $\Gamma_{\pm}^0$ model an external pump, which is used to excite the two level system. In principle, 
it is possible to achieve $\Gamma_{+}^0=\Gamma_{-}^0$. We would like to have a small decay rate $\Gamma_{+,n\rightarrow-,n}^{\rm decay}$
and, therefore, the energy splitting $\Delta E$ should not be resonant with a multiple of $\omega_L$.

Dissipation in the oscillator is described by 
\begin{eqnarray}
L_{\rm diss}\rho
&=&
\frac{\kappa}{2} 
\left(2a\rho a^{\dag}-a^{\dag}a\rho-\rho a^{\dag}a\right).
\end{eqnarray}
Here, the Hamiltonian $H_{\rm sys}$ and the Lindblad operator do not mix the 
off-diagonal and diagonal components of the reduced density matrix. 
Therefore, the system is described by the diagonal components and evolves stochastically. 
We discuss the distribution probability of photons, 
\begin{eqnarray}
\rho_n=\sum_{\sigma=\uparrow,\downarrow} 
\langle \sigma,n| \rho  |\sigma, n\rangle \, , 
\end{eqnarray}
and average number of photons 
$\langle n \rangle = \sum_n n \rho_n$.

 We want to derive an effective equation for the 
 probability distribution of the number of photons in the
 oscillator $\rho_n=\sum_{\sigma}\rho_{\sigma,n}$. We do this by
 tracing out the degrees of freedom
 of the two-level system in the equation of motion given by Eq. (\ref{eq:masterrhoincoherent}).
  Using the relation $\rho_n=\rho_{\uparrow,n}+\rho_{\downarrow,n}$ 
   and the assumption that the time scales of the two-level system are faster than the time scales of the
   oscillator 
   we can form a closed set of equations
   \begin{eqnarray}\label{eq:rhoclosed}
 &  &   \frac{d}{dt}\left(\begin{array}{c}
     \rho_{\uparrow,n-1}\\
     \rho_{\downarrow,n}
    \end{array}
     \right) \\
   & &=\!\!\left(
     \begin{array}{cc}
       -\Gamma_{+}-\!\!\Gamma_{-}-\!\!\Gamma_{+,n-1}^{s}  &                        \Gamma_{-,n}^{s} \\
                              \Gamma_{+,n-1}^{s}  & -\Gamma_{+}-\!\!\Gamma_{-}-\!\!\Gamma_{-,n}^{s} \\
     \end{array}\right)\!\!
     \left(\!\!\!\begin{array}{c}
     \rho_{\uparrow,n-1}\\
     \rho_{\downarrow,n}
    \end{array}\!\!\!\right)\nonumber \\
    & & +\left(\begin{array}{c}
      \Gamma_{+} \rho_{n-1}\\
      \Gamma_{-} \rho_{n}
      \end{array}\right)\, .\nonumber
      \end{eqnarray}
    This set of equations
   can be solved in the stationary limit and we get an equation for the effect of the artificial atom on the
   oscillator
   \begin{eqnarray}\label{eq:effectivn}
    \dot{\rho}_n &=& \gamma_{n}^{+}\rho_{n-1}
             -\left(\gamma_{n+1}^{+}+\gamma_n^{-}+\kappa n\right)\rho_{n}\\
          &     & +\left(\gamma_{n+1}^{-}+\kappa (n+1)\right)\rho_{n+1}\, ,\nonumber
   \end{eqnarray}
 with
 \begin{eqnarray}
  \gamma_{n}^{+}\!\! & = &\!\!
  \frac{\Gamma_+\Gamma_{+,n-1}^s}{\Gamma_+ + \Gamma_- + \Gamma_{+,n-1}^s + \Gamma_{-,n}^s}
  \nonumber \\
  \gamma_n^{-}  \!\! & = &\!\!
   \frac{\Gamma_-\Gamma_{-,n}^s}{\Gamma_+ + \Gamma_- + \Gamma_{+,n-1}^s + \Gamma_{-,n}^s}\, .
  \nonumber \\
 \end{eqnarray}
It is now relatively simple to write down the solution for the stationary density matrix, which is given by
\begin{equation}
 \rho_n=\rho_0 \prod_{m=1}^n\frac{\gamma_{m}^{+}}{\gamma_m^{-}+\kappa m}
\end{equation}
with the normalization factor  $\rho_0$.
In the case of lasing, the distribution function is peaked around the average photon number $\langle a^{\dag}a \rangle = \langle n\rangle$.

 We can derive a good approximation for the average photon number using
\begin{equation}\label{eq_to_find_photon_number}
 \frac{\gamma_{\langle n \rangle}^{+}}{\gamma_{\langle n\rangle} ^{-}+\kappa \langle n \rangle}\approx 1\, .
\end{equation}
This condition can be solved in two regimes which we will discuss next. These two regions are mostly defined 
by the relation between the photon creation/absorption rates $\Gamma_{\pm,n}$ and the broadening $\Gamma$, 
which we introduced in Eq. (\ref{eq_Boradened_spectral_function}).
The broadening can be caused by many effects, like the inherent width of the spectral 
peak around $\omega_L$. However, we also get an additional broadening from e.g. the pumping 
 $\Gamma_{\pm}$ \cite{lasing_with_strong_noise_marthaler}. Therefore, we can always assume $\Gamma\gtrsim (\Gamma_{+}+\Gamma_{-})/2$.

\subsection{Small photon numbers}

We will now consider a regime where the photon number remains relatively small, which means that $\Gamma_T\ll \Gamma_{+}+\Gamma_{-}$.
We use the connection between the broadened spectral function and the pumping rates and assume that the broadening is mostly the result of the pumping.
This gives us $\Gamma_+=\Gamma_-=\Gamma$.
To be more precise, we also assume that the resonance condition $\delta\omega=\omega_L$ is exactly fulfilled.
At low temperatures this allows us to use the following approximations
\begin{eqnarray}
 \gamma_n^+ &=& \frac{1}{2}g^2 n \sin^2(\theta) S_P(\omega_L)-\frac{g^4 n^2 S_P^2(\omega_L)\sin^4(\theta)}{4\Gamma}\nonumber\\
 \gamma_n^- &=& 0~.
\end{eqnarray}
Now it is relatively simple to solve the condition $\gamma_{\langle n \rangle}^{+}/\kappa \langle n \rangle=1$.
From this we get the average photon number
\begin{equation}
 \langle n \rangle \approx \frac{2\Gamma}{g^2\sin^2\theta S_{P}(\omega_L)}-\frac{4\Gamma\kappa}{g^4\sin^2\theta S_{P}(\omega_L)}.
\end{equation}
We can further approximate this solution by assuming that only the resonant peak contributes to $S_{P}(\omega_L)$,
\begin{equation}
 S_{P}(\omega_L)\approx \frac{4\epsilon_C \cos^2\theta}{\Gamma\omega_L}e^{-\frac{4\epsilon_C \cos^2\theta}{\omega_L}},
\end{equation}
which gives us
\begin{eqnarray}
 \langle n \rangle & \approx &\frac{\Gamma^2\omega_L}{2\epsilon_C g^2\sin^2\theta\cos^2\theta}e^{\frac{4\epsilon_C \cos^2\theta}{\omega_L}}\\
                   &         & \times \left(1-\frac{\Gamma\omega_L\kappa}{2 g^2 \epsilon_C \sin^2\theta\cos^2\theta}e^{\frac{4\epsilon_C \cos^2\theta}{\omega_L}}\right)~.\nonumber
\end{eqnarray}

\subsection{Large photon numbers }

In this section we will calculate the photon number in the limit $\Gamma_T\gg \Gamma_{+}+\Gamma_-$.
In this case we can write 
\begin{eqnarray}
  \gamma_{n}^{+}\!\! & = &\!\!
  \frac{\Gamma_+\Gamma_{+,n-1}^s}{ \Gamma_{+,n-1}^s + \Gamma_{-,n}^s}
  \nonumber \\
  \gamma_n^{-}  \!\! & = &\!\!
   \frac{\Gamma_-\Gamma_{-,n}^s}{\Gamma_{+,n-1}^s + \Gamma_{-,n}^s}
  \nonumber \\
 \end{eqnarray}
and we can directly find an effective equation  for the number of photons from Eq. (\ref{eq_to_find_photon_number})
\begin{equation}
 \Gamma_{+,\langle n\rangle-1}^s=
 \frac{\Gamma_-+\kappa \langle n\rangle}{\Gamma_+ - \kappa \langle n\rangle }\Gamma_{-,\langle n\rangle}~.
\end{equation}
To find a more explicit result, we need to simplify the photon creation rates. To do this, we assume we are at small temperatures,
$\omega_L\gg k_B T$, which means $\eta_+\approx 0$, that the resonance condition $\delta\omega=\omega_L$ is met, and $\Gamma_T\ll \omega_L$.
Using these conditions we find
\begin{eqnarray}
 \Gamma_{+,n-1}^{s}&\approx& 2\Gamma-\frac{2\Gamma^3}{g^2\sin^2\theta}\frac{e^{\eta_-}}{\eta_-}\frac{1}{n}\\
 \Gamma_{-,n}^{s}&\approx& \frac{2 g^4 \sin^4\theta}{\Gamma\omega_L^2}f(\eta_-)n^2 \nonumber\\
 f(\eta_-) &=&\eta_- e^{-2\eta_-}\sum_{n'}\frac{\eta_-^{n'}}{n'!}\frac{1}{(n'+1)^2}~. \nonumber
\end{eqnarray}
For the photon creation rates, we want the first term to dominate, which gives us a condition for the coupling strength $g$,
\begin{equation}\label{eq_Condition_to_neglect_escond_term_in_photon_pump_rate}
\frac{\Gamma^2}{g^2\sin^2\theta}\frac{e^{\eta_-}}{\eta_-}\frac{1}{\langle n\rangle}\ll 1.
\end{equation}
This means that in principle we would like the effective coupling between the two level system $g\sin\theta\sqrt{\langle n\rangle}$ and the cavity to be similar to the 
width of the dissipative environment $\Gamma$.
 If the condition (\ref{eq_Condition_to_neglect_escond_term_in_photon_pump_rate})
is valid we find,
\begin{equation}
 \frac{\Gamma^2\omega_L^2}{g^4 \sin^4\theta f(\eta_-)}=\frac{\Gamma_-+\kappa \langle n\rangle}{\Gamma_+ - \kappa \langle n\rangle}\, \langle n\rangle^2.
\end{equation}
For this to have a solution with a large photon number $\langle n \rangle$ we of course need $\Gamma_+ \gg \kappa$. 
Additonally it seems beneficial if $g\ll \omega_L $. As we have seen in condition (\ref{eq_Condition_to_neglect_escond_term_in_photon_pump_rate}),
the coupling $g$ should not be too small. Yet at the same time, $\omega_L$ is relevant for the asymmetry between photon creation and absorption and therefore large $\omega_L$
is an important ingredient. 

\section{Conclusion}
We have discussed lasing in the model of a two-level system coupled to 
a lasing cavity and a dissipative environment. The dissipative environment has a spectral density with a maximum 
at the frequency $\omega_L$, while the lasing cavity has the frequency $\Omega$. Lasing can be generated if we are close to the
resonance condition $\omega_L+\Omega=\Delta E$, where $\Delta E$ is the energy splitting of the two level system. We see that this allows for 
many combinations of $\omega_L$ and $\Omega$ and therefore it allows for lasing at wide range of frequencies. An important feature is that, while 
we operate at small temperatures as compared to the characterisitc energies of the dissipative environment,  $k_B T\ll \omega_L$, the frequency $\Omega$
can be much smaller than temperature.

An interesting regime where these results could be relevant, is for quantum dots in the optical domain with a 
cavity at THz-frequencies. This would allow for   $k_B T\ll \omega_L$ even at room temperatures, and therefore it would allow for 
a room temperature laser in the THz regime.

\section*{Acknowledgements}
The study was supported by a research fellowship
within the project "Enhancing Educational Potential
of Nicolaus Copernicus University in the Disciplines
of Mathematical and Natural Sciences" (project no.
POKL.04.01.01-00-081/10.).

\end{document}